\documentclass[12pt]{article}

\usepackage{amsmath,amssymb,cite}

\def\beq{\begin{equation}}
\def\eqn#1{\beq\label{#1}}
\def\eeq{\end{equation}}

\def\bb {\begin {eqnarray}}
\def\eqnn#1{\bb\label{#1}}
\def\ee {\end {eqnarray}}

\def\bbz{\mathbb{Z}}%{Z\!\!\!Z}
\def\bbc{\mathbb{C}}%{I\!\!\!\!C}}
\def\bac{\bbc} %{{C\kern-5.5pt I}
\def\bbr{\mathbb{R}}%{{I\!\!R}}
\def\bbn{\mathbb{N}}%{I\!\!N}

\def\spa{{\spadesuit}}

\def\nn{\nonumber}
\def\nt{\noindent}

 \def\nd{\end{document}}

\def\nha{{\textstyle{9\over2}}}

\def\llr{\longrightarrow}
\def\({\left(}
\def\){\right)}

\def\lra{\longrightarrow}
\def\llra{\longleftrightarrow}

\def\ha{{\textstyle{1\over2}}}

\def\trha{{\textstyle{3\over2}}}

  \def\tV{{\tilde V}}

\def\r{\rho}

\def\bbr{{I\!\!R}}
\def\bbn{I\!\!N}
\def\a{\alpha}
\def\b{\beta}

\def\vr{\vert}

\def\L{\Lambda}

\def\rank{{\rm rank}}

\def\riga{-\kern-4pt - \kern-4pt -}
\font\fat=cmsy10 scaled\magstep5

\def\Bbullet{\raise-3pt\hbox{\fat\char"0F}}

\def\ca{{\cal A}}  \def\cc{{\cal C}}
\def\cd{{\cal D}} \def\ce{{\cal E}} \def\cf{{\cal F}}
\def\cg{{\cal G}} \def\ch{{\cal H}} 
 \def\ck{{\cal K}} 
\def\cm{{\cal M}} \def\cn{{\cal N}} 
\def\cp{{\cal P}}  
 \def\ct{{\cal T}}

\def\ido{intertwining differential operator}
\def\idos{intertwining differential operators}

 \def\ha{{\textstyle{\frac{1}{2}}}}

\def\ca{{\cal A}}
\def\nn{\nonumber}

\def\nt{\noindent}

\def\lra{\longleftrightarrow}

\input epsf.tex
\newcount\figno
\def\fig#1#2#3{
\par\begingroup\parindent=0pt\leftskip=1cm\rightskip=1cm\parindent=0pt
\baselineskip=11pt \global\advance\figno by 1 %\midinsert
\epsfxsize=#3 \centerline{\epsfbox{#2}} \vskip 12pt
#1\par
\endgroup\par}
\def\figlabel#1{\xdef#1{\the\figno}}
\def\encadremath#1{\vbox{\hrule\hbox{\vrule\kern8pt\vbox{\kern8pt
\hbox{$\displaystyle #1$}\kern8pt} \kern8pt\vrule}\hrule}}
\begin{document}

 \textheight=24cm
  \textwidth=16.5cm
  \topmargin=-1.5cm
  \oddsidemargin=-0.25cm

\begin{center}

{\LARGE {\bf Invariant Differential Operators for Non-Compact Lie
Groups:\\[2pt]
the Reduced  SU(3,3) Multiplets    \footnote{Plenary talk at the
International Workshop 'Supersymmetries and Quantum Symmetries',
Dubna, July 29 - August 3, 2013.}}}

 \vspace{10mm}

{\bf \large V.K. Dobrev}

\vskip 5mm

\emph{Institute for Nuclear
Research and Nuclear Energy, \\
 Bulgarian Academy of Sciences,\\ 72
Tsarigradsko Chaussee, 1784 Sofia, Bulgaria} 
%dobrev@inrne.bas.bg}

\end{center}

\vskip 4mm

\begin{abstract}
In the present paper we continue the project of systematic
construction of invariant differential operators on the example of
the non-compact  algebras  $su(n,n)$. Earlier were given
 the main multiplets of indecomposable elementary
representations for $n\leq 4$, and the reduced ones for $n=2$. Here
we give all reduced multiplets containing physically relevant
representations including the minimal ones for the algebra
$su(3,3)$. Due to the recently established parabolic relations the
results are valid also for the algebra $sl(6,\mathbb{R})$ with
suitably chosen maximal parabolic subalgebra.
\end{abstract}

\section{Introduction}

Invariant differential operators   play very important role in the
description of physical symmetries.
In a recent paper \cite{Dobinv} we started the systematic explicit
construction of invariant differential operators. We gave an
explicit description of the building blocks, namely, the parabolic
subgroups and subalgebras from which the necessary representations
are induced. Thus we have set the stage for study of different
non-compact groups.

 In the present paper we  focus on the  algebra  ~$su(3,3)$.
 The algebras $su(n,n)$ belong to a
narrow class of algebras, which we call 'conformal Lie algebras',
which have very similar properties to the canonical conformal
algebras of  Minkowski space-time. This class was identified from
our point of view in \cite{Dobeseven}. The same class was identified
independently from different considerations and under different
names in \cite{Guna,Mackder}.

 This paper is a   sequel of \cite{Dobsunn}, and due to the lack of space we refer to
it and to  \cite{Dobparab} for motivations and extensive list of
literature on the subject.

\section{Preliminaries}

 Let $G$ be a semisimple non-compact Lie group, and $K$ a
maximal compact subgroup of $G$. Then we have an Iwasawa
decomposition ~$G=KA_0N_0$, where ~$A_0$~ is abelian simply
connected vector subgroup of ~$G$, ~$N_0$~ is a nilpotent simply
connected subgroup of ~$G$~ preserved by the action of ~$A_0$.
Further, let $M_0$ be the centralizer of $A_0$ in $K$. Then the
subgroup ~$P_0 ~=~ M_0 A_0 N_0$~ is a minimal parabolic subgroup of
$G$. A parabolic subgroup ~$P ~=~ M A N$~ is any subgroup of $G$
  which contains a minimal parabolic subgroup.

The importance of the parabolic subgroups comes from the fact that
the representations induced from them generate all (admissible)
irreducible representations of $G$ \cite{Lan,Zhea,KnZu}.

Let ~$\nu$~ be a (non-unitary) character of ~$A$, ~$\nu\in\ca^*$,
let ~$\mu$~ fix an irreducible representation ~$D^\mu$~ of ~$M$~ on
a vector space ~$V_\mu\,$.

 We call the induced
representation ~$\chi =$ Ind$^G_{P}(\mu\otimes\nu \otimes 1)$~ an
~{\it elementary representation} of $G$ \cite{DMPPT}.   Their spaces of functions are:
\eqn{fun} \cc_\chi ~=~ \{ \cf \in C^\infty(G,V_\mu) ~ \vr ~ \cf
(gman) ~=~ e^{-\nu(H)} \cdot D^\mu(m^{-1})\, \cf (g) \} \eeq where
~$a= \exp(H)\in A$, ~$H\in\ca\,$, ~$m\in M$, ~$n\in N$. The
representation action is the $left$ regular action: \eqn{lrr}
(\ct^\chi(g)\cf) (g') ~=~ \cf (g^{-1}g') ~, \quad g,g'\in G\ .\eeq

For our purposes we need to restrict to ~{\it maximal}~ parabolic
subgroups ~$P$, so that $\rank\,A=1$. Thus, for our  representations
the character ~$\nu$~ is parameterized by a real number ~$d$,
called the conformal weight or energy.

An important ingredient in our considerations are the ~{\it
highest/lowest weight representations}~ of ~$\cg$. These can be
realized as (factor-modules of) Verma modules ~$V^\L$~ over
~$\cg^\bac$, where ~$\L\in (\ch^\bac)^*$, ~$\ch^\bac$ is a Cartan
subalgebra of ~$\cg^\bac$, weight ~$\L = \L(\chi)$~ is determined
uniquely from $\chi$ \cite{Har,Dob}.

Actually, since our ERs will be induced from finite-dimensional
representations of ~$\cm$~ (or their limits)  the Verma modules are
always reducible. Thus, it is more convenient to use ~{\it
generalized Verma modules} ~$\tV^\L$~ such that the role of the
highest/lowest weight vector $v_0$ is taken by the
 space ~$V_\mu\,v_0\,$. For the generalized
Verma modules (GVMs) the reducibility is controlled only by the
value of the conformal weight $d$. Relatedly, for the \idos{} only
the reducibility w.r.t. non-compact roots is essential.

One main ingredient of our approach is as follows. We group the
(reducible) ERs with the same Casimirs in sets called ~{\it
multiplets} \cite{Dobmul,Dob}. The multiplet corresponding to fixed
values of the Casimirs may be depicted as a connected graph, the
vertices of which correspond to the reducible ERs and the lines
between the vertices correspond to intertwining operators. The
explicit parametrization of the multiplets and of their ERs is
important for understanding of the situation.

In fact, the multiplets contain explicitly all the data necessary to
construct the \idos{}. Actually, the data for each \ido{} consists
of the pair ~$(\b,m)$, where $\b$ is a (non-compact) positive root
of ~$\cg^\bac$, ~$m\in\bbn$, such that the BGG \cite{BGG} Verma
module reducibility condition (for highest weight modules) is
fulfilled: \eqn{bggr} (\L+\r, \b^\vee ) ~=~ m \ , \quad \b^\vee
\equiv 2 \b /(\b,\b) \ .\eeq When \eqref{bggr} holds then the Verma
module with shifted weight ~$V^{\L-m\b}$ (or ~$\tV^{\L-m\b}$ ~ for
GVM and $\b$ non-compact) is embedded in the Verma module ~$V^{\L}$
(or ~$\tV^{\L}$). This embedding is realized by a singular vector
~$v_s$~ determined by a polynomial ~$\cp_{m,\b}(\cg^-)$~ in the
universal enveloping algebra ~$(U(\cg_-))\ v_0\,$, ~$\cg^-$~ is the
subalgebra of ~$\cg^\bac$ generated by the negative root generators
\cite{Dix}.
More explicitly, \cite{Dob}, ~$v^s_{m,\b} = \cp^m_{\b}\, v_0$ (or ~$v^s_{m,\b} = \cp^m_{\b}\, V_\mu\,v_0$ for GVMs).
  Then there exists \cite{Dob} an \ido{} \eqn{lido}
\cd^m_{\b} ~:~ \cc_{\chi(\L)} ~\llr ~ \cc_{\chi(\L-m\b)} \eeq given
explicitly by: \eqn{mido}\cd^m_{\b} ~=~ \cp^m_{\b}(\widehat{\cg^-})
\eeq where ~$\widehat{\cg^-}$~ denotes the $right$ action on the
functions ~$\cf$, cf. \eqref{fun}.

\section{The non-compact Lie algebra $su(3,3)$}

\nt Let ~$\cg ~=~ su(3,3)$.   This algebra
has discrete series representations and highest/lowest weight
representations since the maximal
compact subalgebra is ~$\ck \cong u(1)\oplus su(3)\oplus su(3)$.

We choose a ~{\it maximal} parabolic ~$\cp=\cm\ca\cn$~ such that
~$\ca\cong so(1,1)$, ~$\cm = ~sl(3,\bbc)_\bbr\,$. We note also that ~$\ck^\bac \cong
u(1)^\bac \oplus sl(3,\bbc) \oplus sl(3,\bbc) \cong \cm^\bac \oplus
\ca^\bac$. Thus,  the factor ~$\cm$~ has the same
finite-dimensional (nonunitary) representations as the
finite-dimensional (unitary) representations of  the semi-simple
subalgebra of   ~$\ck$.

We label   the signature of the ERs of $\cg$   as follows:
\eqn{sgnd}  \chi ~=~ \{\, n_1\,, n_{2}\,, n_{4}\,, n_{5}\,;\, c\, \} \ , \qquad n_j \in \bbz_+\ , \quad c =
d- \nha \eeq where the last entry of ~$\chi$~ labels the characters of
$\ca\,$, and the first $4$ entries are labels of the
finite-dimensional nonunitary irreps of $\cm$ when all $n_j>0$  or
limits of the latter when some $n_j=0$.

Below we shall use the following conjugation on the
finite-dimensional entries of the signature: \eqn{conu}
(n_1,n_{2},n_{4},n_{5})^* ~\doteq~
(n_{4},n_{5},n_1,n_{2})  \ . \eeq

The ERs in the multiplet are related also by intertwining integral
  operators  introduced in \cite{KnSt}. These operators are defined
for any ER,   the general action
being: \eqnn{knast}  && G_{KS} ~:~ \cc_\chi ~ \llr ~ \cc_{\chi'} \
,\cr &&\chi ~=~ \{\, n_1,n_{2},n_{4},n_{5}
\,;\, c\, \} \ , \qquad \chi' ~=~ \{\,
(n_1,n_{2},n_{4},n_{5})^* \,;\, -c\, \} . \ee

For the classification of the multiplets we shall need one more conjugation
for the entries of the ~$\cm$~ representations:
\eqn{conut} (n_1,n_{2},n_{4},n_{5})^\spa ~\doteq~
(n_{5},n_4,n_{2},n_{1}) \ . \eeq

Further, we need the root system of the complexification
~$\cg^\bac = sl(6,\bbc)$~.  The positive roots in terms of the
simple roots are given standardly as:
\eqnn{sunnpos} \a_{ij} ~&=&~ \a_i + \cdots + \a_{j}\ , \quad 1 \leq i < j \leq
5 \ , \cr \a_{jj} ~&=&~ \a_j \ , \quad 1 \leq j \leq 5 \ee
From these the compact roots are those that form (by
restriction) the root system of the semisimple part of ~$\ck^\bac$,
the rest are noncompact, i.e., \eqnn{sunncnc}   {\rm noncompact:}&~~~ \a_{ij}\ , \quad 1
\leq i \leq 3 \ , ~~ \quad 3 \leq j \leq 5 \ .
  \ee

Further, we give the correspondence between the signatures $\chi$
and the highest weight $\L$. The connection is through the Dynkin
labels:    \eqn{dynk} m_i ~\equiv~ (\L+\r,\a^\vee_i) ~=~ (\L+\r,
\a_i )\ , \quad i=1,\ldots,5, \eeq where ~$\L = \L(\chi)$, ~$\r$
is half the sum of the positive roots of ~$\cg^\bac$.  The explicit
connection is: \eqn{rela} n_i = m_i \ , \quad  c ~=~ - m_\r
 ~=~  -\,\ha(   m_1+ 2m_{2} + 3m_3 + 2m_{4} + m_{5})
 \eeq

We shall use also   the so-called Harish-Chandra parameters:
\eqn{dynhc} m_{jk} \equiv (\L+\r, \a_{jk} ) ~=~ m_j + \cdots + m_k\ , ~~j<k\ ,
\qquad
m_{jj} \equiv m_j\ .\eeq

Finally, we remind that according to  \cite{Dobparab} the above considerations
for ~$su(3,3)$~ are applicable also for the algebra ~$sl(6,\bbr)$~
with parabolic ~$\cm$-factor ~$sl(3,\bbr)\oplus sl(3,\bbr)$.

\section{Multiplets of SU(3,3)}

\subsection{Main multiplets}

There are two types of multiplets: main and reduced.
The multiplets of the main type are in 1-to-1 correspondence
with the finite-dimensional irreps of ~$su(3,3)$, i.e., they are
labelled by  the five  positive Dynkin labels    ~$m_i\in\bbn$.
   In \cite{Dobsunn}  we have given
explicitly the main multiplets for ~$n=2,3,4$, and the reduced for $n=2$.

 A main multiplet contains 20 ERs/GVMs whose signatures can be
given in the following pair-wise manner \cite{Dobsunn}:
\eqnn{tabltri}  \chi_0^\pm
   &=&    \{  (
 m_1,
 m_2,
 m_4,
 m_5)^\pm ; \pm m_\r  \} \nn\\
\chi_a^\pm      &=&      \{  (
 m_1,
 m_{23},
 m_{34},
 m_5)^\pm ; \pm (m_\r  -  m_3)  \} \nn\\
\chi_b^\pm    &=&    \{  (
 m_{12},
 m_{3},
 m_{24},
 m_5)^\pm ; \pm  (m_\r  -  m_{23})  \} \nn\\
\chi_{b'}^\pm    &=&    \{  (
 m_{1},
 m_{24},
 m_{3},
 m_{45})^\pm ; \pm   (m_\r  -  m_{34})  \} \nn\\
\chi_c^\pm      &=&      \{  (
 m_{2},
 m_{3},
 m_{14},
 m_5)^\pm ; \pm (m_\r  -  m_{13})  \} \nn\\
\chi_{c'}^\pm          &=&          \{  (
 m_{12},
 m_{34},
 m_{23},
 m_{45})^\pm ; \pm (m_\r  -  m_{24})  \} \nn\\
\chi_{c''}^\pm      &=&      \{  (
 m_{1},
 m_{25},
 m_{3},
 m_{4})^\pm ; \pm (m_\r  -  m_{35})  \} \nn\\
\chi_d^\pm        &=&        \{  (
 m_{2},
 m_{34},
 m_{13},
 m_{45})^\pm ; \pm (m_\r  -  m_{14})  \} \nn\\
\chi_{d'}^\pm        &=&        \{  (
 m_{12},
 m_{35},
 m_{23},
 m_{4})^\pm ; \pm (m_\r  -  m_{25})  \} \nn\\
\chi_e^\pm      &=&      \{  (
 m_{2},
 m_{35},
 m_{13},
 m_{4})^\pm ; \pm  (m_\r  -  m_{15})    \}   \ee
 where ~$(k_1,k_2,k_3,k_4)^- = (k_1,k_2,k_3,k_4)$,
~$(k_1,k_2,k_3,k_4)^+ = (k_1,k_2,k_3,k_4)^*$.
 They are given explicitly in Fig.~1 (first in \cite{Dobsunn}).
The pairs ~$\L^\pm$~ are symmetric w.r.t. to the bullet in the
middle of the figure - this represents the Weyl symmetry realized by
the Knapp-Stein operators \eqref{knast}:~ $G_{KS} ~:~ \cc_{\chi^\mp}
\lra \cc_{\chi^\pm}\,$.

Matters are arranged so that in every multiplet only the ER with
signature ~$\chi_0^-$~ contains a finite-dimensional nonunitary
subrepresentation in  a finite-dimensional subspace ~$\ce$. The
latter corresponds to the finite-dimensional   irrep of ~$su(3,3)$~ with
signature ~$\{ m_1\,, \ldots\,, m_5 \}$.   The subspace ~$\ce$~ is annihilated by the
operator ~$G^+\,$,\ and is the image of the operator ~$G^-\,$. The
subspace ~$\ce$~ is annihilated also by the \ido{} acting from
~$\chi^-_0$~ to ~$\chi^-_a\,$.
 When all ~$m_i=1$~ then ~$\dim\,\ce = 1$, and in that case
~$\ce$~ is also the trivial one-dimensional UIR of the whole algebra
~$\cg$. Furthermore in that case the conformal weight is zero:
~$d=\nha+c=\nha-\ha(m_1+2m_2+3m_3+2m_4+m_5)_{\vert_{m_i=1}}=0$.

In the conjugate ER ~$\chi_0^+$~ there is a unitary
subrepresentation in  an infinite-dimen\-sional subspace  $\cd$. It
is annihilated by the operator  $G^- $,\ and is the image of the
operator  $G^+ $.

 All the above is valid also for the algebra ~$sl(6,\bbr)$, cf.  \cite{Dobparab}.
 However, the latter algebra does not have
 discrete series representations. On the other hand the algebra
 ~$su(3,3)$~ has discrete series representations and
 furthermore highest/lowest weight series representations.

Thus, for  ~$su(3,3)$~ in every multiplet only the ER with signature
~$\chi_0^+$~ contains a holomorphic discrete series representation.
The ER ~$\chi_0^+\,$~ contains also the conjugate
anti-holomorphic discrete series. The direct sum of the holomorphic
and the antiholomorphic representations is the
invariant subspace ~$\cd$~ of the ER ~$\chi_0^+\,$.
Note that the corresponding lowest weight GVM
is infinitesimally equivalent only to the holomorphic discrete
series, while the conjugate highest weight GVM is infinitesimally
equivalent to the anti-holomorphic discrete series.

In Fig.~1 and below we use the notation: ~$\L^\pm = \L(\chi^\pm)$.
Each \ido\ is represented by an
arrow accompanied by a symbol ~$i_{jk}$~ encoding the root
~$\a_{jk}$~ and the number $m_{\a_{jk}}$ which is involved in
the BGG criterion.  This notation is used to save space, but it can
be used due to the fact that only \idos\ which are
non-composite are displayed, and that the data ~$\b,m_\b\,$, which
is involved in the embedding ~$V^\L \lra V^{\L-m_\b,\b}$~ turns out
to involve only the ~$m_i$~ corresponding to simple roots, i.e., for
each $\b,m_\b$ there exists ~$i = i(\b,m_\b,\L)\in \{
1,\ldots,5\}$, such that ~$m_\b=m_i\,$. Hence the data
~$\a_{jk}\,$,~$m_{\a_{jk}}$~ is represented by ~$i_{jk}$~
on the arrows.

\subsection{Reduced multiplets}

  There are five types of reduced
multiplets, $R_a\,$, $a=1,\ldots,5$,  which may be obtained from the
main multiplet by setting formally ~$m_a=0$. Multiplets of type
$R_4\,$, $R_5\,$, are conjugate  to the
multiplets of type $R_2\,$, $R_1\,$, resp., as follows.
First we make the conjugation on the roots and exchange all indices: ~$1 \llra 5$, ~$2 \llra 4$.
With this operation we obtain the diagrams of the conjugated cases from one another.
For the entries of the ~$\cm$~ representation we have further to employ the conjugation
\eqref{conut}. Then we obtain the signatures of the conjugated cases from one another.
Thus, we give explicitly only first three types.

The reduced multiplets of type $R_3$ contain 14 ERs/GVMs whose
signatures can be given in the following pair-wise manner:
\eqnn{tabltrtr} \chi_0^\pm ~&=&~ \{\, (
 m_1,
 m_2,
 m_4,
 m_5)^\pm\,;\,\pm m_\r  \,\}  \\
\chi_b^\pm ~&=&~ \{\, (
 m_{12},
 0,
 m_{24},
 m_5)^\pm\,;\,\pm  (m_\r  -  m_{2}) \,\} \cr
\chi_{b'}^\pm ~&=&~ \{\, (
 m_{1},
 m_{24},
 0,
 m_{45})^\pm\,;\,\pm (m_\r  -  m_{4}) \,\} \cr
\chi_c^\pm ~&=&~ \{\, (
 m_{2},
 0,
 m_{14},
 m_5)^\pm\,;\,\pm (m_\r  -  m_{12}) \,\} \cr
\chi_{c''}^\pm ~&=&~ \{\, (
 m_{1},
 m_{25},
 0,
 m_{4})^\pm\,;\,\pm (m_\r  -  m_{45}) \,\} \cr
\chi_d^\pm ~&=&~ \{\, (
 m_{2},
 m_{4},
 m_{12},
 m_{45})^\pm\,;\,\pm (m_\r  -  m_{12,4}) = \mp (m_\r  -  m_{2,45}) \,\} \cr
\chi_e^\pm ~&=&~ \{\, (
 m_{2},
 m_{45},
 m_{12},
 m_{4})^\pm\,;\,\pm (m_\r  -  m_{2,4}) = \pm \ha (m_1+m_5) \,\} \ ,\nn\ee
 here ~$m_\r = \ha (m_1 + 2m_2 + 2m_4 + m_5)$.
 These multiplets are given in Fig. 2.
 They may be called the main type of reduced multiplets since
here in ~$\chi_0^+$~ are contained the limits of the
(anti)holomorphic  discrete series.

The reduced multiplets of type $R_2$ contain 14 ERs/GVMs whose
signatures can be given in the following pair-wise manner:
\eqnn{tabltrtw}
\chi_0^\pm ~&=&~ \{\, (
 m_1,
 0,
 m_4,
 m_5)^\pm\,;\,\pm m_\r \,\} \\
\chi_b^\pm ~&=&~ \{\, (
 m_{1},
 m_{3},
 m_{34},
 m_5)^\pm\,;\,\pm (m_\r - m_3) \,\} \cr
\chi_c^\pm ~&=&~ \{\, (
 0,
 m_{3},
 m_{1,34},
 m_5)^\pm\,;\,\pm (m_\r-m_{1,3})\,\} \cr
\chi_{c'}^\pm ~&=&~ \{\, (
 m_{1},
 m_{34},
 m_{3},
 m_{45})^\pm\,;\,\pm (m_\r - m_{34})\,\} \cr
\chi_d^\pm ~&=&~ \{\, (
 0,
 m_{34},
 m_{1,3},
 m_{45})^\pm\,;\,\pm (m_{\r}-m_{1,34})\,\} \cr
\chi_{d'}^\pm ~&=&~ \{\, (
 m_{1},
 m_{35},
 m_{3},
 m_{4})^\pm\,;\,\pm (m_{\r}-m_{35} )\,\} \cr
\chi_e^\pm ~&=&~ \{\, (
 0,
 m_{35},
 m_{1,3},
 m_{4})^\pm\,;\,\pm (m_\r - m_{1,35}) \,\} \ , \nn\ee
 here ~$m_\r = \ha (m_1 + 3m_3 + 2m_4 + m_5)$.
 These multiplets are given in Fig. 3.

%\medskip

The reduced multiplets of type $R_1$ contain 14 ERs/GVMs whose
signatures can be given in the following pair-wise manner:
\eqnn{tabltron}
\chi_0^\pm ~&=&~ \{\, (
 0,
 m_2,
 m_4,
 m_5)^\pm\,;\,\pm m_\r  \,\} \\
\chi_a^\pm ~&=&~ \{\, (
 0,
 m_{23},
 m_{34},
 m_5)^\pm\,;\,\pm (m_\r - m_3  )\,\} \cr
\chi_b^\pm ~&=&~ \{\, (
 m_{2},
 m_{3},
 m_{24},
 m_5)^\pm\,;\,\pm (m_\r - m_{23}) \,\} \cr
\chi_{b'}^\pm ~&=&~ \{\, (
 0,
 m_{24},
 m_{3},
 m_{45})^\pm\,;\,\pm (m_\r - m_{34})\,\} \cr
\chi_{c}^\pm ~&=&~ \{\, (
 0,
 m_{25},
 m_{3},
 m_{4})^\pm\,;\,\pm (m_\r - m_{35}) \,\} \cr
\chi_d^\pm ~&=&~ \{\, (
 m_{2},
 m_{34},
 m_{23},
 m_{45})^\pm\,;\,\pm (m_\r - m_{24}) \,\} \cr
\chi_e^\pm ~&=&~ \{\, (
 m_{2},
 m_{35},
 m_{23},
 m_{4})^\pm\,;\,\pm (m_\r - m_{25}) \,\}\ , \nn\ee
  here ~$m_\r = \ha (2m_2 + 3m_3 + 2m_4 + m_5)$.
  These multiplets are given in Fig. 4.

\subsection{Further reduction of multiplets}

There are further reductions of the multiplets denoted by
~$R^3_{ab}\,$, $a,b=1,\ldots,5$, $a< b$, which may be obtained from
the main multiplet by setting formally ~$m_a=m_b=0$. From these ten
reductions four (for $(a,b)=(1,2),(2,3),(3,4),(4,5)$) do not contain
representations of physical interest, i.e., induced from
finite-dimensional irreps of the ~$\cm$~ subalgebra. From the others
~$R^3_{35}$~ and ~$R^3_{25}$~ are conjugated to ~$R^3_{13}$~ and
~$R^3_{14}\,$, resp., as explained above. Thus, we present
explicitly only four types of multiplets.

The reduced multiplets of type $R^3_{13}$ contain 10 ERs/GVMs whose
signatures can be given in the following pair-wise manner:
\eqnn{tabltrona}
\chi_a^\pm ~&=&~ \{\, (
 0,
 m_2,
 m_4,
 m_5)^\pm\,;\,\pm m_\r  \,\} \\
\chi_b^\pm ~&=&~ \{\, (
 m_{2},
0,
 m_{2,4},
 m_5)^\pm\,;\,\pm (m_\r - m_{2}) \,\} \cr
\chi_{b'}^\pm ~&=&~ \{\, (
 0,
 m_{2,4},
 0,
 m_{45})^\pm\,;\,\pm (m_\r - m_{4})\,\} \cr
\chi_{c}^\pm ~&=&~ \{\, (
 0,
 m_{2,45},
 0,
 m_{4})^\pm\,;\,\pm (m_\r - m_{45}) \,\} \cr
\chi_d^\pm ~&=&~ \{\, (
 m_{2},
 m_{4},
 m_{2},
 m_{45})^\pm\,;\,\pm (m_\r - m_{2,4}) = \pm\ha m_5\,\}
 \ ,\nn \ee
  here ~$m_\r = m_2 + m_4 + \ha m_5$.
The multiplets are given in Fig. 5.
Note that the differential operator (of order $m_5$) from ~$\chi_d^-$~
 to ~$\chi_d^+$~ is a degeneration of an integral Knapp-Stein operator.

The reduced multiplets of type $R^3_{14}$ contain 10 ERs/GVMs whose
signatures can be given in the following pair-wise manner:
\eqnn{tabltronb}
\chi_0^\pm ~&=&~ \{\, (
 0,
 m_2,
 0,
 m_5)^\pm\,;\,\pm m_\r  \,\} \\
\chi_a^\pm ~&=&~ \{\, (
 0,
 m_{23},
 m_{3},
 m_5)^\pm\,;\,\pm (m_\r - m_3  )\,\} \cr
\chi_b^\pm ~&=&~ \{\, (
 m_{2},
 m_{3},
 m_{23},
 m_5)^\pm\,;\,\pm (m_\r - m_{23}) \,\} \cr
 \chi_{c}^\pm ~&=&~ \{\, (
 0,
 m_{23,5},
 m_{3},
 0)^\pm\,;\,\pm (m_\r - m_{3,5}) \,\} \cr
 \chi_d^\pm ~&=&~ \{\, (
 m_{2},
 m_{3,5},
 m_{23},
 0)^\pm\,;\,\pm (m_\r - m_{23,5})= \pm \ha (m_3 - m_5) \,\}\ ,\nn \ee
  here ~$m_\r = \ha (2m_2 + 3m_3 + m_5)$.
The multiplets are given in Fig. 6.

The reduced multiplets of type $R^3_{15}$ contain 10 ERs/GVMs whose
signatures can be given in the following pair-wise manner:
\eqnn{tabltronc}
\chi_0^\pm ~&=&~ \{\, (
 0,
 m_2,
 m_4,
 0)^\pm\,;\,\pm m_\r  \,\} \\
\chi_a^\pm ~&=&~ \{\, (
 0,
 m_{23},
 m_{34},
 0)^\pm\,;\,\pm (m_\r - m_3  )\,\} \cr
\chi_b^\pm ~&=&~ \{\, (
 m_{2},
 m_{3},
 m_{24},
 0)^\pm\,;\,\pm (m_\r - m_{23}) \,\} \cr
\chi_{b'}^\pm ~&=&~ \{\, (
 0,
 m_{24},
 m_{3},
 m_{4})^\pm\,;\,\pm (m_\r - m_{34})\,\} \cr
 \chi_d^\pm ~&=&~ \{\, (
 m_{2},
 m_{34},
 m_{23},
 m_{4})^\pm\,;\,\pm (m_\r - m_{24})=\pm \ha m_3 \,\}
 \ , \nn\ee
  here ~$m_\r = m_2 + \trha m_3 + m_4$.
The multiplets are given in Fig. 7.
Here  the differential operator (of order $m_3$) from ~$\chi_d^-$~
 to ~$\chi_d^+$~  is a degeneration  of an integral Knapp-Stein operator.

The reduced multiplets of type $R^3_{24}$ contain 10 ERs/GVMs whose
signatures can be given in the following pair-wise manner:
\eqnn{tabltrtwa}
\chi_0^\pm ~&=&~ \{\, (
 m_1,
 0,
 0,
 m_5)^\pm\,;\,\pm m_\r \,\} \\
\chi_b^\pm ~&=&~ \{\, (
 m_{1},
 m_{3},
 m_{3},
 m_5)^\pm\,;\,\pm (m_\r - m_3) \,\} \cr
\chi_c^\pm ~&=&~ \{\, (
 0,
 m_{3},
 m_{13},
 m_5)^\pm\,;\,\pm (m_\r-m_{1,3})\,\} \cr
 \chi_{d}^\pm ~&=&~ \{\, (
 m_{1},
 m_{35},
 m_{3},
 0)^\pm\,;\,\pm (m_{\r}-m_{3,5} )\,\} \cr
\chi_e^\pm ~&=&~ \{\, (
 0,
 m_{35},
 m_{13},
 0)^\pm\,;\,\pm (m_\r - m_{1,3,5})=\pm  \ha (m_3 - m_1 - m_5) \,\} \ , \nn\ee
 here ~$m_\r = \ha (m_1 + 3m_3 + m_5)$.
The multiplets are given in Fig. 8.

\subsection{Last reduction of multiplets}

There are further reductions of the multiplets - triple and quadruple,
but only one triple reduction contains representations of physical interest.
Namely, this is the multiplet ~$R^3_{135}\,$,  which
may be obtained from the main multiplet by setting formally ~$m_1=m_3=m_5=0$.
It contains 7 ERs/GVMs whose
signatures can be given in the following  manner:
\eqnn{tabltrona}
\chi_a^\pm ~&=&~ \{\, (
 0,
 m_2,
 m_4,
 0)^\pm\,;\,\pm m_\r =\pm  m_{2,4} \,\} \\
\chi_b^\pm ~&=&~ \{\, (
 m_{2},
0,
 m_{2,4},
 0)^\pm\,;\,\pm m_4  \,\} \cr
\chi_{b'}^\pm ~&=&~ \{\, (
 0,
 m_{2,4},
 0,
 m_{4})^\pm\,;\,\pm m_2  \,\} \cr
\chi_d ~&=&~ \{\, (
 m_{2},
 m_{4},
 m_{2},
 m_{4})\,;\, 0  \,\} \nn \ee
The multiplets are given in Fig. 9.
The representation ~$\chi_d$~ is a singlet, not in a pair, since it has zero weight ~$c$,
and the ~$\cm$~ entries are self-conjugate under \eqref{conu}.  It is placed in the middle
of the figure as the bullet. That ER contains the ~{\it minimal irreps}~ of ~$SU(3,3)$~ characterized by two positive integers
which are denoted in this context as ~$m_2\,,m_4\,$. Each such irrep is the kernel of the two
invariant differential operators ~$\cd^{m_2}_{14}$~ and ~$\cd^{m_4}_{25}$, which are of order ~$m_2\,$, ~$m_4\,$, resp.,
and correspond to the noncompact roots ~$\a_{14}\,$, ~$\a_{25}\,$, resp., cf. \eqref{mido}.

\begin{footnotesize}

\section*{Acknowledgments}

The author would like to thank the Organizers for the kind
invitation to speak at the International Workshop 'Supersymmetries
and Quantum Symmetries', Dubna, July 29 - August 3, 2013. The author
has received partial support from COST action MP-1210.

\end{footnotesize}

%\np

%\np

\vspace{5mm}

\fig{}{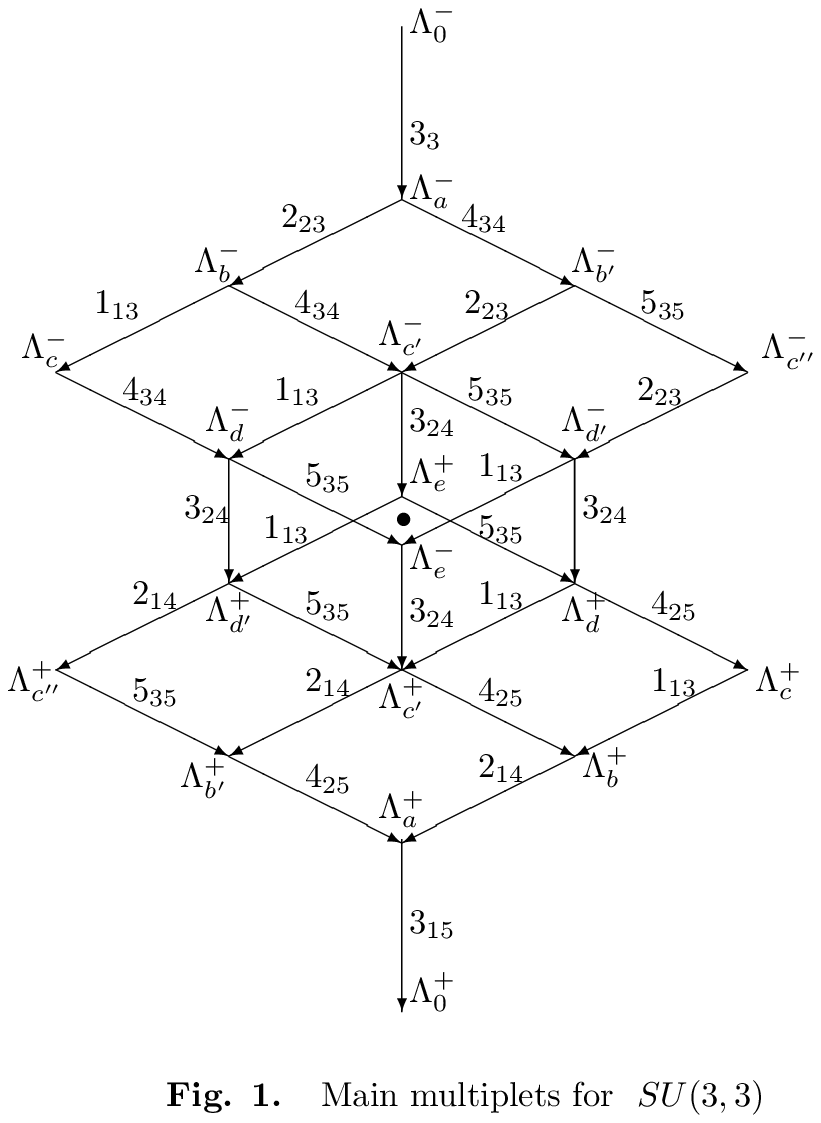}{6cm}
\fig{}{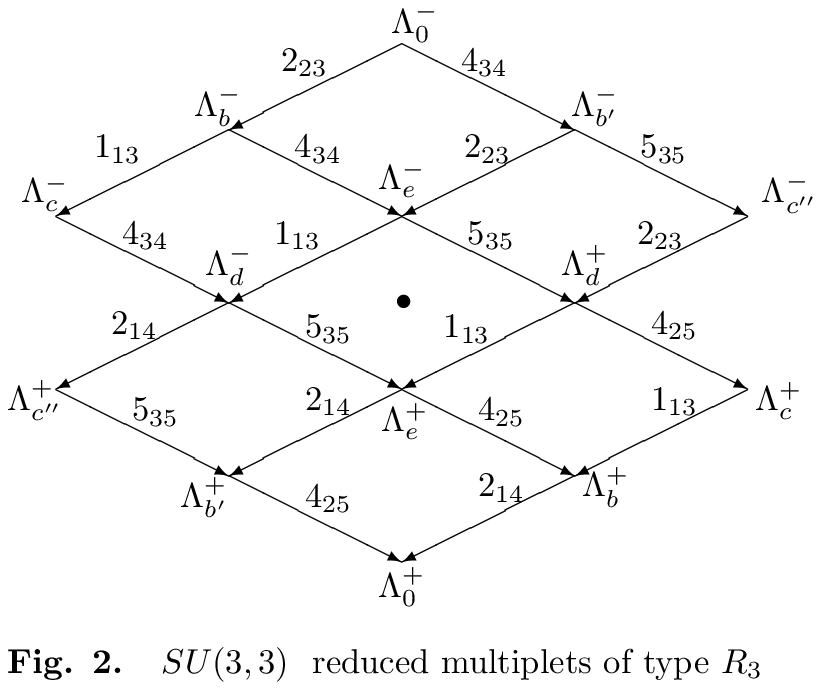}{7cm}
\fig{}{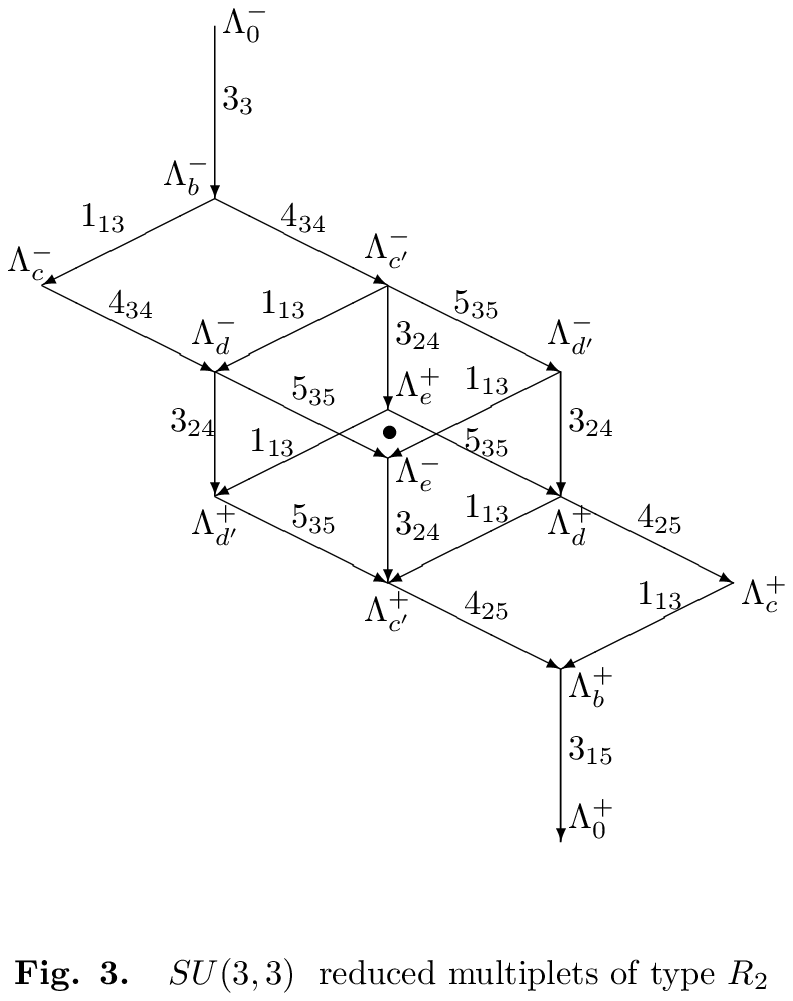}{7cm}
\fig{}{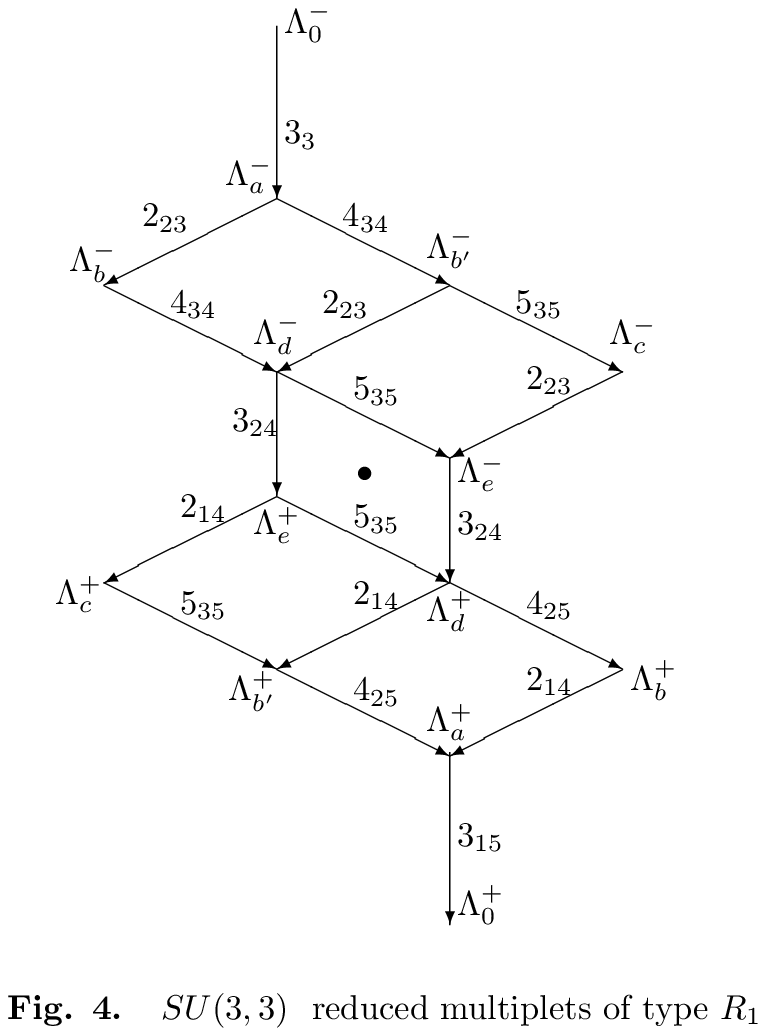}{7cm}

\fig{}{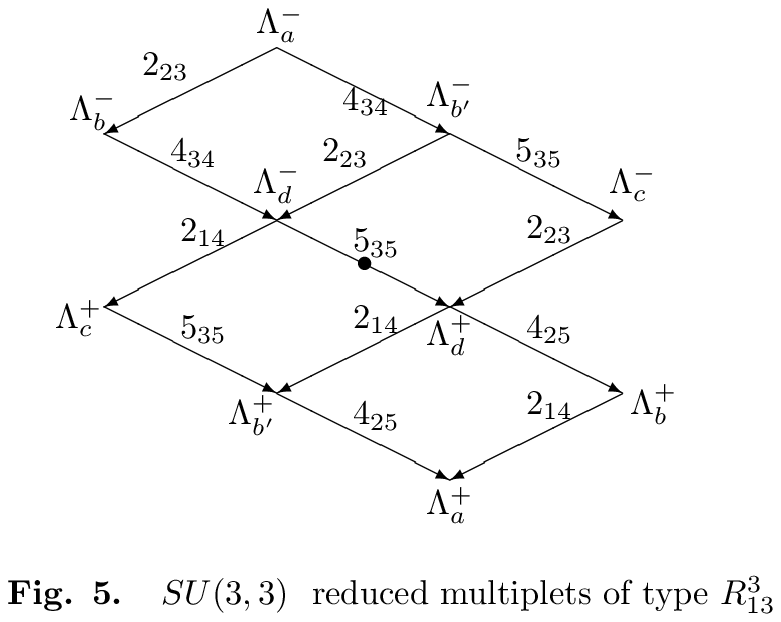}{8cm}
\fig{}{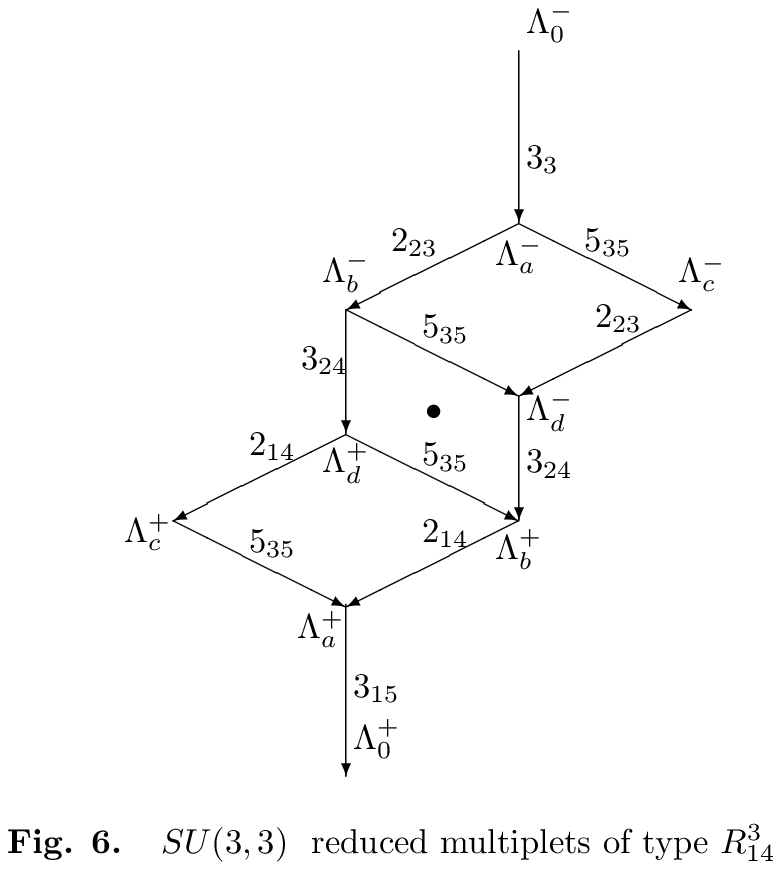}{8cm}
\fig{}{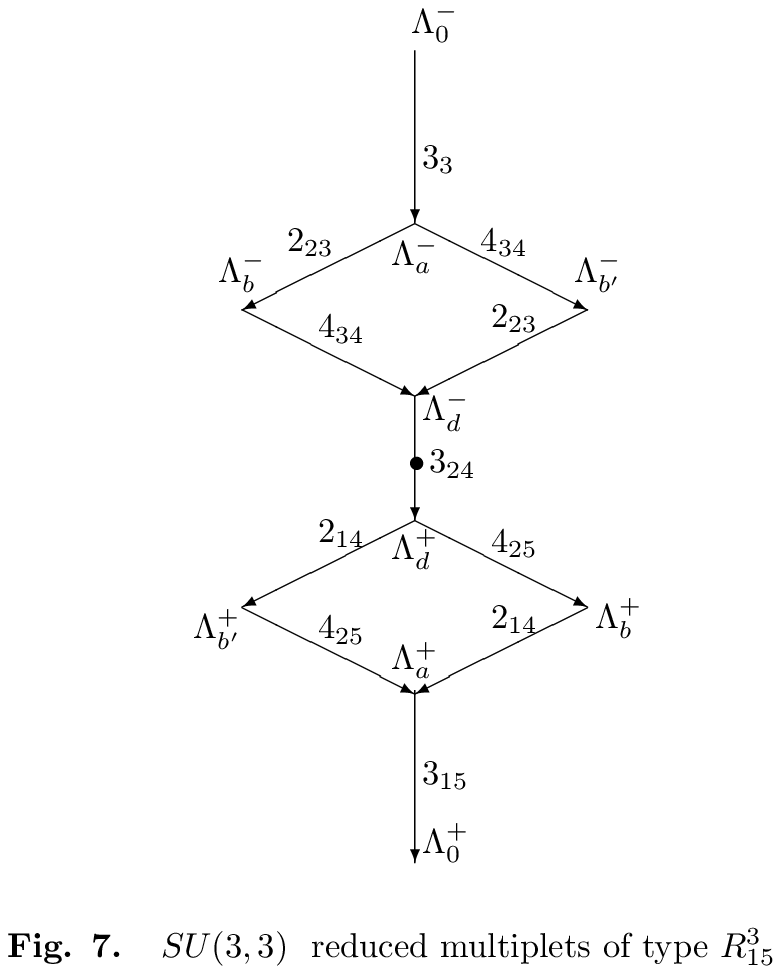}{8cm}
\fig{}{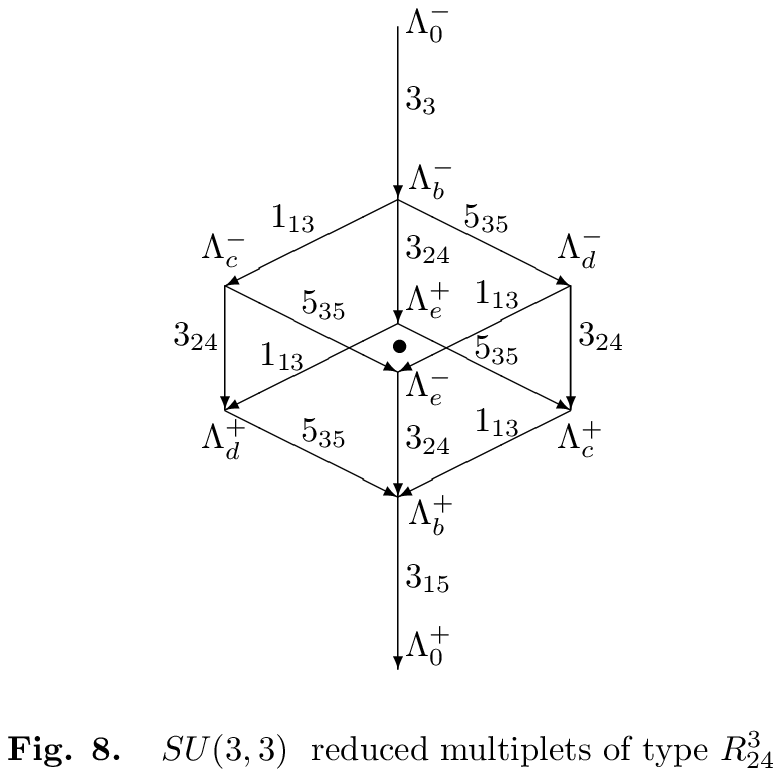}{8cm}
\fig{}{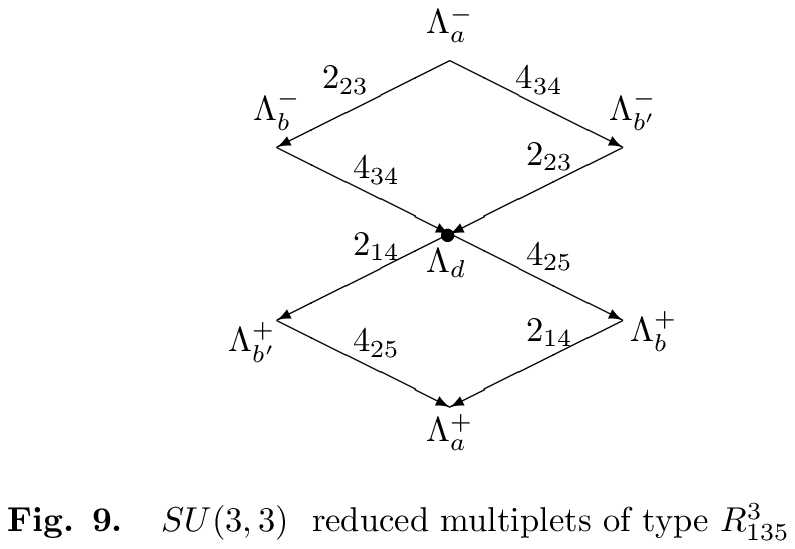}{8cm}

\end{document}